\documentclass[twocolumn,pdflatex,sn-nature]{sn-jnl}

\usepackage{geometry}
\usepackage{graphicx}%
\usepackage{multirow}%
\usepackage{amsmath,amssymb,amsfonts}%
\usepackage{amsthm}%
\usepackage{mathrsfs}%
\usepackage[title]{appendix}%
\usepackage{xcolor}%
\usepackage{colortbl}
\usepackage{textcomp}%
\usepackage{manyfoot}%
\usepackage{booktabs}%
\usepackage{algorithm}%
\usepackage{algorithmicx}%
\usepackage{algpseudocode}%
\usepackage{listings}%

\theoremstyle{thmstyleone}%
\theoremstyle{thmstyletwo}%

\theoremstyle{thmstylethree}%

\begin{document}

\title[Article Title]{A Pre-trained Framework for Multilingual Brain Decoding Using Non-invasive Recordings}

\author[1,2]{\fnm{Yi} \sur{Guo}}
\author[1,2]{\fnm{Yihang} \sur{Dong}}
\author[3]{\fnm{Michael Kwok-Po} \sur{Ng}}
\author*[1,2]{\fnm{Shuqiang} \sur{Wang}}\email{sq.wang@siat.ac.cn}



\affil[1]{\orgdiv{Shenzhen Institute of Advanced Technology}, \orgname{Chinese Academy of Sciences}, \orgaddress{\city{Shenzhen}, \country{China}}}

\affil[2]{\orgname{University of Chinese Academy of Sciences}, \orgaddress{\city{Beijing}, \country{China}}}
\affil[3]{\orgname{Hong Kong Baptist University}, \orgaddress{\city{Hong Kong}, \country{China}}}

\abstract{Brain-computer interfaces (BCIs) with speech decoding from brain recordings have broad application potential in fields such as clinical rehabilitation and cognitive neuroscience. However, current decoding methods remain limited to single-language, single-subject, and single neuroimaging modality settings, restricting their clinical applicability and generalizability. Here we propose a joint multilingual, multi-subject and multimodal decoding framework. It maps diverse brain recordings into a unified semantic space defined by a pre-trained multilingual model (PMM), enabling decoding across multiple languages, multiple subjects and multiple neuroimaging modalities. The proposed framework is validated using non-invasive brain recordings from 159 participants across four languages. Experimental results show that it exhibits strong generalization across multilingual, multi-subject, and multimodal settings. More importantly, the proposed framework can promote linguistic fairness, which is vital for underrepresented languages in BCI applications. The unified semantic space enables cross-lingual mapping enhancement, allowing the framework to boost the decoding performance of underrepresented languages, thereby promoting linguistic fairness. Overall, the proposed framework establishes a new potential paradigm for brain decoding, opening new paths for broader applications of BCI.}

\maketitle

\section*{Introduction}\label{sec1}

\begin{figure*}[htbp]
    \centering
    \includegraphics[width=\textwidth]{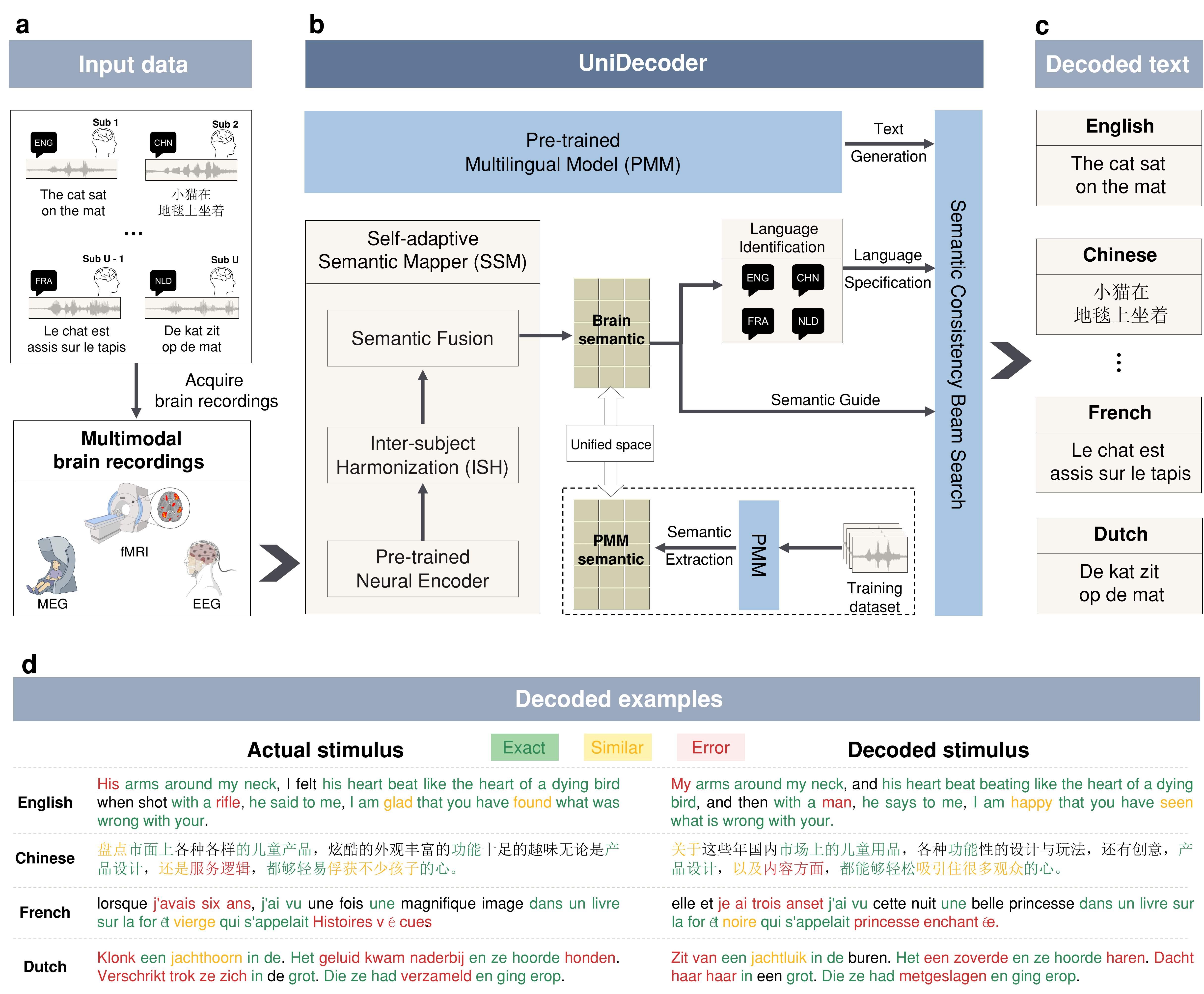}
    \caption{\textbf{Schematic of the proposed brain decoding framework.} \textbf{a}, Brain recordings were collected from multiple subjects while they listened to narratives in different languages, with data acquired across multiple neuroimaging modalities. \textbf{b}, Schematic of the UniDecoder framework. SSM maps diverse brain recordings into a unified semantic space defined by the PMM. This yields a brain semantic representation. The stimulus language is identified from the brain semantic representation to specify the target language for text generation. Semantic consistency beam search integrates the brain semantic representation, the identified language, and PMM to generate the decoded text. During training, SSM is optimized to align the brain semantic representation with the PMM semantic representation extracted from stimulus text in the training dataset. \textbf{c}, Decoded text generated by UniDecoder. \textbf{d}, Examples of multilingual decoding segments. Decoding examples showing pairs of actual stimulus text and corresponding decoded output for English, Chinese, French, and Dutch separately. Brain responses were recorded while subjects listened to test narratives not used in training.} %
    \label{fig:1}
\end{figure*}

BCIs with speech decoding show great promise both in restoring communication for patients with aphasia \cite{RN18,RN38,RN1} and in advancing understanding of the neural mechanisms underlying human language \cite{RN72,RN37,us+1}. While invasive brain recordings using implanted electrodes have enabled accurate speech decoding \cite{RN57,RN49,RN14}, their broader adoption is hindered by the surgical risks of implantation, suboptimal long-term reliability, and substantial associated costs \cite{RN36,RN73}. Non-invasive brain recording methods, such as functional magnetic resonance imaging (fMRI), magnetoencephalography (MEG), and electroencephalography (EEG), offer safer and more accessible alternatives \cite{us+7}. Decoding speech from non-invasive brain recordings is showing increasing promise and has attracted increasing attention \cite{RN7,RN2,RN50}.

Although non-invasive brain decoding methods have advanced considerably in recent years, their practical utility remains limited by persistent challenges. First, the linguistic scope of brain decoding research remains narrow. Most existing studies focus on a single language—predominantly English—restricting the applicability of BCIs in multilingual contexts \cite{RN81}. Achieving global accessibility requires decoding methods that generalize across languages and promote linguistic fairness for underrepresented languages \cite{RN70}. However, substantial differences in phonemic inventories and syntactic structures across languages pose significant challenges for developing robust multilingual brain decoders. Although recent studies based on invasive recordings have demonstrated bilingual decoding in constrained vocabularies \cite{RN15}, open-vocabulary multilingual decoding using non-invasive methods remains an unresolved challenge. Second, variability in brain activity across individuals limits the generalization of decoding methods \cite{RN80,us+4,us+6}. Most current approaches rely on subject-specific models to address inter-individual variability, but this reliance compromises generalization and hinders the broader applicability of BCIs \cite{RN7}. Finally, non-invasive brain recordings are intrinsically constrained by limited spatial and temporal resolution, which fundamentally limits decoding performance \cite{RN79,us+3,us+10}. While multimodal integration offers a promising strategy to overcome these limitations and enhance decoding accuracy \cite{RN60,RN61,RN78,us+2,us+9}, existing approaches remain confined to single-modality decoding \cite{RN52,RN7,RN2}.

In recent years, PMMs have been shown to capture high-level semantic information across languages and exhibit brain-like representational patterns during language processing \cite{RN3,RN4,RN5,RN39}. These properties suggest that PMMs may serve as a bridge for aligning brain activity with semantic representations across different languages. Building on this foundation, this work proposes a new strategy for multilingual brain decoding that constructs a unified semantic space using representations generated by the PMM, enabling brain recordings from different languages to be mapped into the unified semantic space. This unified semantic space offers an alignment mechanism for cross-lingual mapping. It further supports the integration of multimodal neuroimaging recordings at the semantic level \cite{us+11,us+12}, and facilitates cross-subject generalization by aligning brain recordings from different individuals into a shared representational structure.

Building on this strategy, the unified brain decoder (UniDecoder) framework is proposed as a brain decoding approach applicable to multilingual, multi-subject, and multimodal settings. UniDecoder maps diverse brain recordings into a unified semantic space and generates natural language text from the resulting semantic representations (Fig. \ref{fig:1}b). This semantic space is defined by the PMM, which encodes high-level natural language semantics. To map brain recordings into this space, UniDecoder incorporates an Self-adaptive semantic mapper (SSM). This module integrates pre-trained neural encoders for extracting features from different neuroimaging modalities, applies an inter-subject harmonization (ISH) module to align representations across participants, and merges multimodal semantic features into a unified representation. Subsequently, the framework combines the PMM with a semantic consistency-based beam search to generate natural language text from the unified semantic representation. By projecting diverse brain recordings into a unified semantic space, UniDecoder enables decoding under multilingual, multi-subject, and multimodal settings.

We validated the UniDecoder framework on four non-invasive brain recording datasets, including 159 participants across four languages (English, Chinese, French, Dutch) and three neuroimaging modalities (fMRI, MEG, EEG). The framework generalizes across languages, participants, and modalities, successfully reconstructing natural language text that reflects the semantic content encoded in brain activity. SHAP-based analysis \cite{SHAP} revealed similar cortical contribution patterns across languages, suggesting shared semantic mechanisms. Building on this finding, we further demonstrate that the unified semantic space of UniDecoder enables cross-lingual enhancement, allowing the framework to improve decoding performance for underrepresented languages. This multilingual capability reduces data requirements under low-resource language conditions, thereby promoting linguistic fairness in BCI applications.

\section*{Results}\label{sec2}

\subsection*{Generalizable brain decoding across diverse datasets}\label{subsec1}
To evaluate the overall generalizability of UniDecoder, decoding experiments were conducted across four datasets comprising naturalistic auditory stimuli: SMN4Lang \cite{RN10}, LPPC-fMRI \cite{RN8}, Broderick2018 \cite{RN32}, and SparrKULee \cite{RN11}. These datasets were selected to provide diverse experimental conditions for assessing the framework’s ability to reconstruct semantically relevant text from brain activity (see Extended Data Table 1 for dataset details). Decoding performance was evaluated using four standard language similarity metrics: word error rate (WER), BLEU-1 \cite{RN53}, METEOR \cite{RN54}, and BGEScore \cite{RN26}, which capture different aspects of similarity between decoded outputs and reference texts. Specifically, WER, BLEU-1, and METEOR focus on lexical and syntactic correspondence, while BGEScore quantifies sentence-level semantic similarity using embedding-based representations.

\begin{figure*}[htbp]
    \centering
    \includegraphics[width=\textwidth]{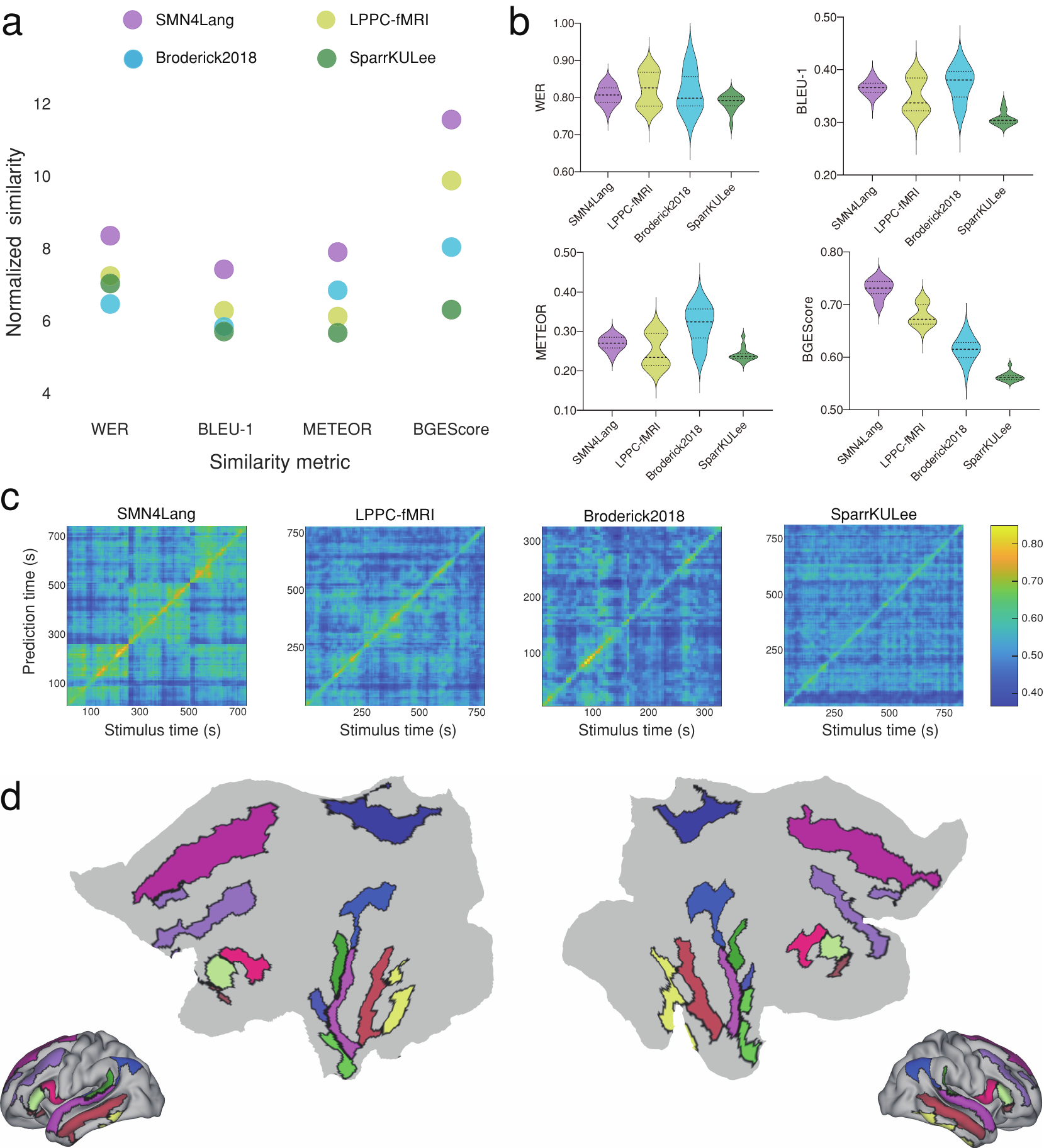}
    \caption{\textbf{Evaluation of brain decoding performance using UniDecoder across four datasets.} \textbf{a}, Comparison of decoding performance across four datasets using language similarity metrics. Scores are normalized using z-scores for dimensionless comparisons (see Methods, Language similarity metrics). \textbf{b}, Violin plots showing the distribution of original language similarity scores across datasets. The width of each violin reflects the distribution of similarity scores. Three horizontal dashed lines indicate the 25\textsuperscript{th} percentile (Q1), 50\textsuperscript{th} percentile (Q2, median), and 75\textsuperscript{th} percentile (Q3), collectively defining the interquartile range (IQR). \textbf{c}, Semantic similarity matrices show the BGEScore between decoded and stimulus-aligned texts over time. Each matrix entry ($i$, $j$) indicates the BGEScore between the decoded text at prediction time $i$ and the reference text at stimulus time $j$, averaged across all subjects.  \textbf{d}, Visualization of language-related brain regions of interest (ROIs), utilized by UniDecoder for brain decoding from fMRI data (see Methods, Data preprocessing). Sample sizes for each dataset are provided in Extended Data Table 1.}
    \label{fig:2}
\end{figure*}

The decoding results are summarized in Fig. \ref{fig:2}, demonstrating that semantically relevant content can be effectively reconstructed from brain recordings across all datasets. Moreover, the decoded outputs partially recover accurate words and syntactic structures (Fig. \ref{fig:1}d). In Fig. \ref{fig:2}a, decoding performance is assessed using the four similarity metrics, with scores computed relative to randomized baselines and normalized for cross-metric comparison. The distribution of similarity scores for each dataset, presented in Fig. \ref{fig:2}b, further demonstrates the consistency of decoding performance. Median WER scores are close to 0.80 across all datasets, while BLEU-1 and METEOR medians cluster around 0.35 and 0.30, respectively. BGEScore medians are highest for SMN4Lang and LPPC-fMRI, both exceeding 0.70. In addition to overall accuracy, temporal alignment between decoded outputs and stimulus sequences is evaluated in Fig. \ref{fig:2}c. The results show that UniDecoder captures the temporal structure of the stimuli, and off-diagonal similarities in the alignment matrix suggesting that contextual semantic information is integrated over short timescales during language processing. Together, these findings confirm that UniDecoder generalizes well across datasets.

\subsection*{Unified representation enables multilingual brain decoding}\label{subsec2}
\begin{figure*}[htbp]
    \centering
    \includegraphics[width=1.0\textwidth]{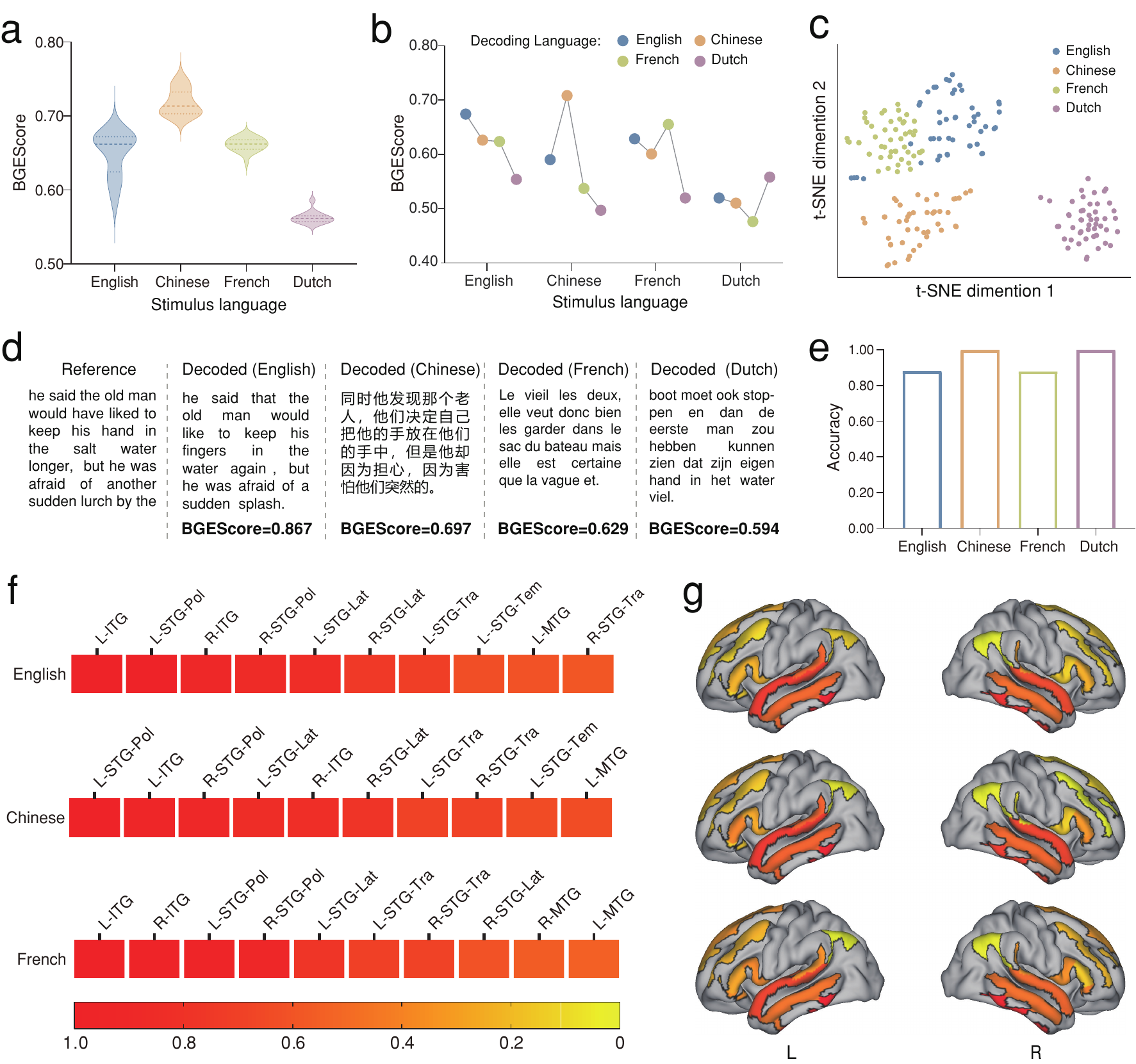}
    \caption{\textbf{Multilingual brain decoding and semantic cortical analysis using UniDecoder.} \textbf{a}, Violin plots showing BGEScore distributions for English, Chinese, French, and Dutch. Scores were pooled across all four datasets. \textbf{b}, Cross-language decoding performance. Each point shows the BGEScore for a given stimulus language (x-axis) and decoding language (color), with higher scores observed when decoding uses the same language as the stimulus. \textbf{c}, t-SNE visualization of the unified semantic representations derived from brain recordings, showing that representations corresponding to different stimulus languages are separated in the semantic space. \textbf{d}, Example text with corresponding decoded outputs across four languages, demonstrating performance degradation when decoding language differs from stimulus language. \textbf{e}, Accuracy of the language identification module applied to the unified semantic representations, showing high performance across all four languages. \textbf{f}, Top-10 cortical regions contributing to semantic processing in English, Chinese, and French are ranked within the language-related brain regions. \textbf{g}, Spatial visualization of cortical contribution patterns on the brain surface, showing the similarity of cortical regions involved in semantic processing across the three languages.}
    \label{fig:3}
\end{figure*}

Multilingual brain decoding experiments are conducted on four datasets, which collectively include brain recordings for English, Chinese, French, and Dutch. As shown in Fig. \ref{fig:3}a, Chinese achieves the highest BGEScore, while French and English show similar performance, and Dutch exhibits the lowest BGEScore among the four languages. The reduced performance for Dutch may be related to the higher tokenization complexity required for Dutch sentences in the PMM (Extended Data Fig. 2). These results demonstrate that the UniDecoder framework enables effective decoding of semantic information from brain recordings across diverse languages.

The preceding results demonstrate that UniDecoder accurately decodes semantic content when the decoding language matches the stimulus language of the brain recordings. However, in practical scenarios where the stimulus language is unknown, selecting an incorrect decoding language can lead to substantial performance degradation. This limitation is confirmed by cross-language decoding experiments, in which mismatches between the decoding language and the stimulus language result in marked decreases in decoding accuracy (Fig. \ref{fig:3}b). A representative example is shown in Fig. \ref{fig:3}d, where the semantic similarity between decoded outputs and reference texts drops sharply under language mismatch conditions. Ideally, an effective decoding framework should be able to identify the intended language directly from brain activity to ensure accurate decoding. To examine whether the learned semantic representations preserve language-specific information, the unified semantic space was visualized using t-SNE (Fig. \ref{fig:3}c). This analysis reveals distinct clustering patterns corresponding to different languages, indicating that language identity is implicitly encoded within the semantic representations. To leverage this information, a language recognition module was introduced to predict the stimulus language from brain activity. As shown in Fig. \ref{fig:3}e, this module achieves high identification accuracy across English, Chinese, French, and Dutch, enabling automatic selection of the appropriate decoding model. These results demonstrate that UniDecoder leverages unified semantic representations to automatically determine and decode the intended language, enabling accurate and automated multilingual brain decoding.

\subsection*{UniDecoder reveal cortical contributions to multilingual brain semantic processing	}\label{subsec3}
Leveraging the capability of UniDecoder to map brain recordings into a unified semantic representation space, cortical contribution patterns during semantic processing across different languages are systematically analyzed. Brain recordings from the LPPC-fMRI dataset \cite{RN8}, acquired while participants listened to narratives in English, Chinese, and French, are used for this experiment. SHAP-based interpretability \cite{SHAP} is applied to quantify the contribution of each cortical region to the transformation of neural activity into semantic representations. As shown in Fig. \ref{fig:3}f, a core set of temporal and inferior frontal regions consistently contributes to brain decoding across all three languages. The left inferior temporal gyrus (L-ITG) and left superior temporal gyrus (L-STG) exhibit prominent and reliable contributions, reflecting their established roles in lexical-semantic integration and higher-order language comprehension \cite{RN6, RN66}. While the overall contribution patterns are largely shared, the L-STG shows relatively stronger involvement in the Chinese condition compared to English and French, which may reflect additional phonological and tonal processing demands specific to tonal languages \cite{RN6}. Spatial distribution maps further indicate bilateral involvement in semantic processing, with highly similar cortical contribution patterns observed across languages, suggesting a common neural basis for semantic representation irrespective of linguistic background (Fig. \ref{fig:3}g).

The observed similarity in cortical contribution patterns across English, Chinese, and French indicates the existence of a shared neural architecture underlying multilingual semantic processing. This convergence highlights the potential to leverage language-invariant brain regions for cross-lingual enhancement in brain decoding. By integrating a unified semantic representation with region-wise contribution analysis, the UniDecoder framework systematically identifies core cortical substrates that support semantic processing across typologically distinct languages. Such an approach not only advances mechanistic understanding but also provides a principled basis for optimizing BCI systems. Focusing signal acquisition and decoding on cortical areas with consistently high contributions may facilitate the development of more efficient and broadly applicable multilingual BCI technologies.

\subsection*{UniDecoder enables multi-subject brain decoding and promotes linguistic fairness }\label{subsec5}
\begin{figure*}[htbp]
    \centering
    \includegraphics[width=\textwidth]{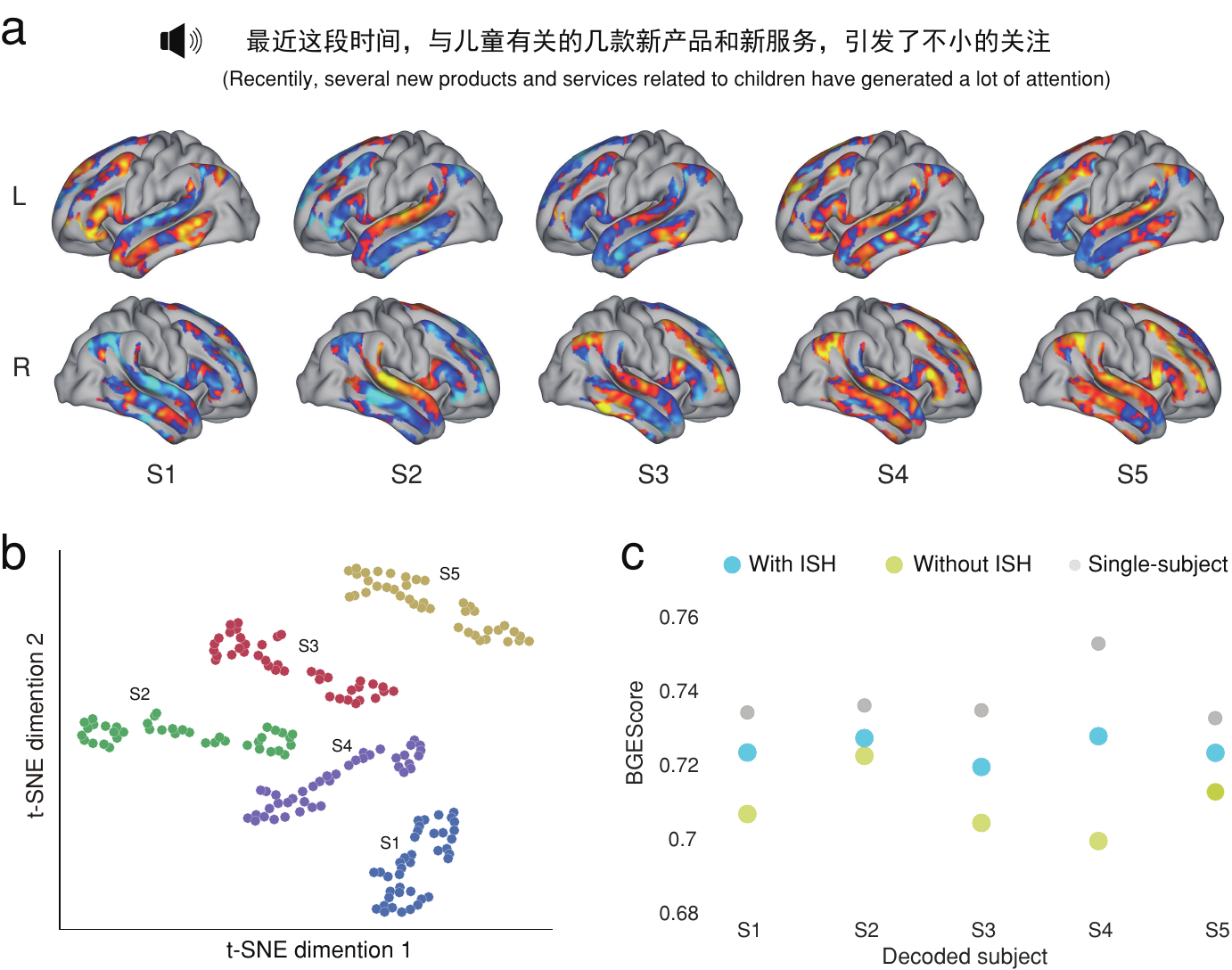}
    \caption{\textbf{Multi-subject brain decoding using UniDecoder.} \textbf{a}, fMRI activation maps from five subjects in the SMN4Lang dataset responding to identical auditory stimuli, visualized within language-related ROIs, showing substantial inter-subject variability. \textbf{b}, t-SNE visualization of brain recordings from five subjects, forming subject-specific clusters in the feature space. \textbf{c}, Evaluation of UniDecoder for multi-subject brain decoding. BGEScore comparison across five subjects under three decoding settings: UniDecoder trained separately for each subject, a multi-subject UniDecoder trained without ISH, and a multi-subject UniDecoder with ISH. UniDecoder with ISH achieves performance close to the single-subject setting, demonstrating its effectiveness in multi-subject decoding.}
    \label{fig:5}
\end{figure*}

Brain activity patterns show notable variability across individuals. For example, identical external stimuli can evoke distinct activation patterns between subjects, as visualized in the SMN4Lang dataset (Fig. \ref{fig:5}a). This inter-subject difference is further illustrated by subject-specific clustering in the t-SNE projection of brain representations (Fig. \ref{fig:5}b). As a consequence, applying a shared decoding model across multiple subjects leads to reduced semantic decoding performance. To address this, an ISH module within the UniDecoder framework was applied to align brain representations across subjects. Compared with models without ISH, decoding semantic similarity was consistently improved, narrowing the gap with subject-specific models (Fig. \ref{fig:5}c). These findings indicate that mitigating individual variability through harmonization enhances multi-subject decoding performance, facilitating broader adoption of BCI technologies.

\begin{figure*}[htbp]
    \centering
    \includegraphics[width=\textwidth]{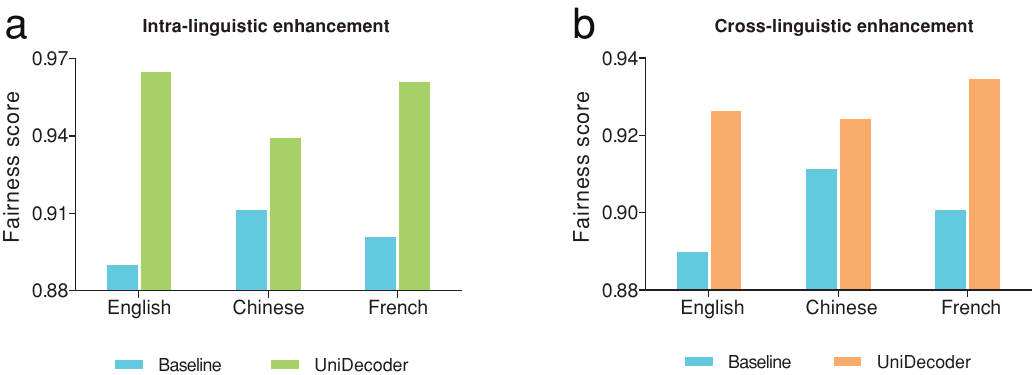}
    \caption{\textbf{Linguistic fairness in brain decoding using UniDecoder.} Data-limited conditions were simulated on the LPPC-fMRI dataset by randomly reducing 60\% of each subject’s training samples, and fairness scores were calculated as the ratio between BGEScore obtained under reduced and full data conditions. \textbf{a}, Intra-linguistic enhancement. For each language, data from other subjects sharing the same language is leveraged to improve decoding under data-limited conditions through intra-lingual mapping enhancement. \textbf{b}, Cross-linguistic enhancement. For each language, data from subjects from other languages is incorporated to support decoding under data-limited conditions through cross-lingual mapping enhancement.}
    \label{fig:6}
\end{figure*}

To address the challenge of decoding performance degradation under data scarcity, we simulated data-limited conditions on the LPPC-fMRI dataset by randomly reducing 60\% of each subject’s training samples. Decoding performance was quantified using the BGEScore. To evaluate decoding generalization under limited data, we defined a fairness score as the ratio between BGEScore obtained under reduced and full data conditions. Intra-linguistic enhancement was first assessed by integrating data from other subjects within the same language, leading to improvements of fairness scores from 0.89 to 0.97 in English, 0.90 to 0.94 in Chinese, and 0.90 to 0.97 in French (Fig. \ref{fig:6}a). These results demonstrate that UniDecoder effectively improves decoding robustness for data-limited individuals within the same language environment. Cross-linguistic enhancement was further evaluated by introducing data from subjects speaking different languages, where the target language was supported by the combined data of the other two languages. This strategy also resulted in consistent gains, with fairness scores increasing to 0.93 in English, 0.92 in Chinese, and 0.94 in French (Fig. \ref{fig:6}b), indicating that UniDecoder can generalize this enhancement across languages through leveraging shared semantic representations. Together, these findings suggest that UniDecoder provides a viable strategy to mitigate disparities in brain decoding performance and enhance lingual fairness, particularly for underrepresented languages.

\begin{figure*}[htbp]
	\centering
	\includegraphics[width=\textwidth]{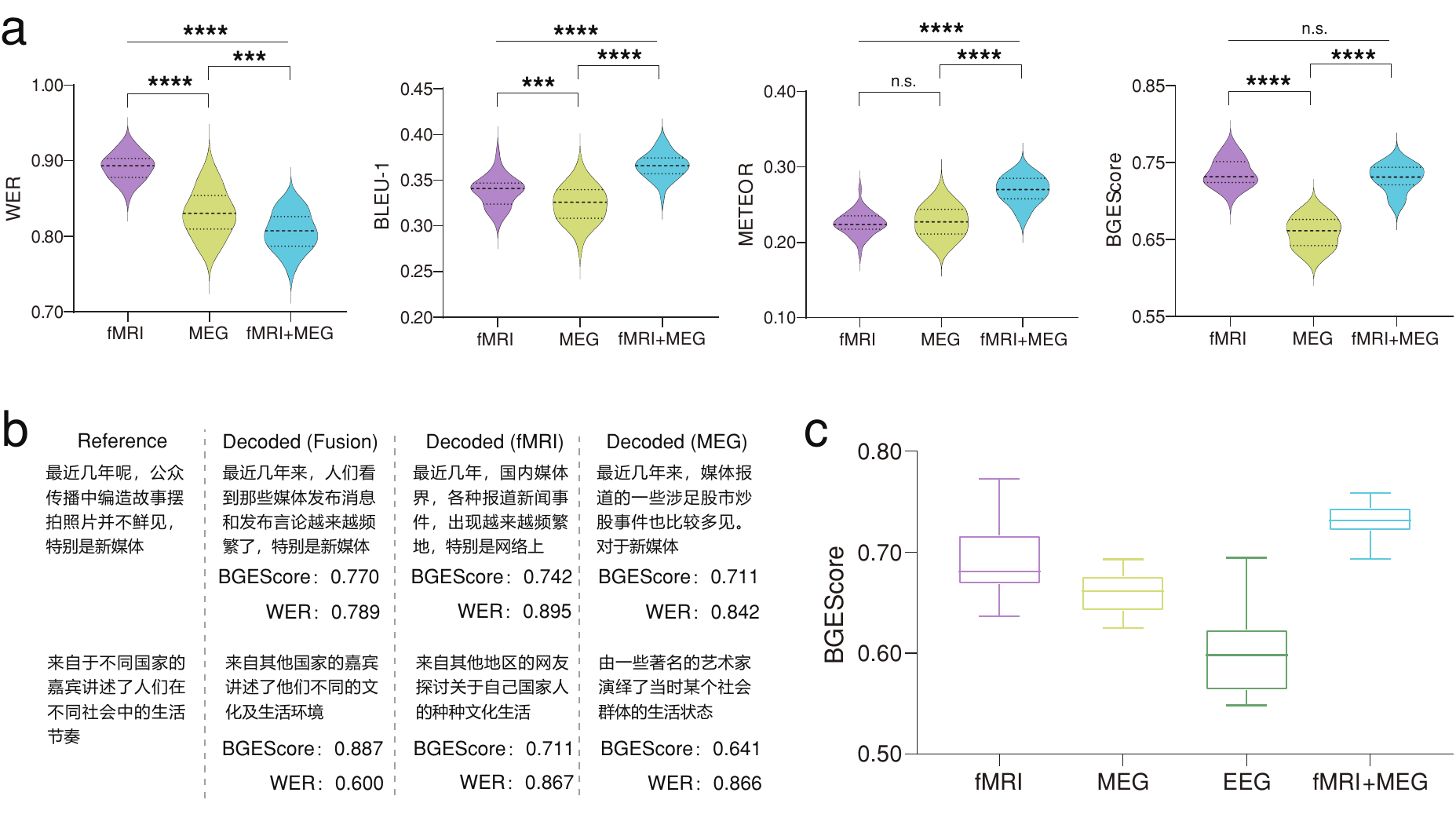}
	\caption{\textbf{Decoding performance of UniDecoder across single and multimodal neuroimaging conditions.} \textbf{a}, Violin plots comparing decoding performance across individual modalities (fMRI, MEG) and the fusion modality (fMRI+MEG) using four linguistic similarity metrics, using all 12 subjects from the SMN4Lang dataset. Each violin shows the score distribution across modalities, with width reflecting the distribution and horizontal lines marking the median and quartiles. Statistical comparisons were performed using repeated-measures one-way ANOVA with Tukey’s post hoc test. *** means \( P < 0.001 \) and **** means \( P < 0.0001 \), and n.s. indicates non-significant differences (\( P > 0.05 \)). \textbf{b}, Representative decoded text examples from individual and fusion modalities, showing that fusion improves overall decoding performance by leveraging complementary strengths from different modalities. \textbf{c}, A box plot showing BGEScore across individual modalities (fMRI, MEG, EEG) and the fusion modality (fMRI+MEG), aggregated across four datasets. Box plots indicate the median (horizontal line), 25\textsuperscript{th} and 75\textsuperscript{th} percentiles (box), and minimum and maximum values (whiskers). Performance improvements with modality combination are consistently observed. For WER, BLEU-1, and METEOR comparisons, see Extended Data Fig. 1.}
	\label{fig:4}
\end{figure*}

\subsection*{Multimodal fusion improves brain decoding performance}\label{subsec4}
To systematically assess the effect of multimodal integration on brain decoding, we compared single-modality and fusion conditions using the UniDecoder framework on the SMN4Lang dataset. As shown in Fig.~\ref{fig:4}a, MEG outperformed fMRI on word-level metrics, achieving significantly better performance on both BLEU-1 and WER (both $P < 0.0001$). In contrast, fMRI yielded higher scores on the semantic-level metric BGEScore, significantly outperforming MEG ($P < 0.0001$), indicating its advantage in capturing global semantic representations. Fusion yielded significantly higher scores than fMRI and MEG in WER ($P < 0.0001$ and $P < 0.001$), and in both BLEU-1 and METEOR (both $P < 0.0001$), demonstrating consistent word-level improvements. For BGEScore, fusion significantly outperformed MEG ($P < 0.0001$), but did not differ significantly from fMRI ($P = 0.2964$). Nevertheless, fusion exhibited a narrower interquartile range than fMRI (0.022 vs. 0.027), suggesting more stable decoding performance. Fig.~\ref{fig:4}b shows a representative decoding example, in which the fusion condition yielded a BGEScore of 0.770 and WER of 0.789, outperforming fMRI (BGEScore 0.742, WER 0.895) and MEG (BGEScore 0.711, WER 0.842). We next compared decoding performance across all four datasets to assess the relative effectiveness of different neuroimaging modalities. As shown in Fig.~\ref{fig:4}c, the fusion condition achieved the highest median BGEScore (0.731), exceeding that of any single-modality configuration. Among individual modalities, fMRI yielded a median score of 0.681, outperforming MEG (0.662) and EEG (0.598), while EEG exhibited the lowest performance overall.

The integration of fMRI and MEG within the UniDecoder framework consistently improved overall decoding performance compared with either modality alone, as evidenced by higher similarity scores and reduced performance variability. These findings demonstrate the benefit of combining complementary spatial and temporal information to enhance decoding accuracy and robustness, and support multimodal fusion as an effective strategy to advance brain decoding.

\section*{Discussion}\label{sec3}
We propose UniDecoder, a brain decoding framework designed to overcome the prevailing limitations of existing methods that remain constrained to single-language, single-subject and single-modality decoding. By mapping diverse brain recordings into a unified semantic space defined by PMM and integrating semantic consistency beam search, UniDecoder enables robust and generalizable decoding of brain recordings into natural language text across multiple languages, neuroimaging modalities and participants. Its effectiveness has been validated using non-invasive recordings from 159 participants encompassing fMRI, MEG and EEG signals across English, Chinese, French and Dutch. In addition to achieving high semantic similarity scores as measured by BGEScore, UniDecoder demonstrates the capacity to generate linguistically faithful outputs that capture the semantic intent, with many decoded sequences also recovering exact characters, words or phrases from the presented stimuli. These findings collectively support UniDecoder as a broadly applicable brain decoding approach, while offering a promising direction for enhancing lingual fairness in BCI applications.

To our knowledge, UniDecoder is the first framework to support multilingual brain decoding using non-invasive brain recordings. Traditional brain decoding methods typically require language-specific models, which substantially increase computational costs and limit scalability in BCI applications. In contrast, UniDecoder projects brain recordings into a unified semantic representation space defined by PMMs, which capture high-level semantic information across languages and exhibit brain-like representational patterns during language processing. This unified space provides an effective alignment mechanism that enables multilingual decoding without the need for language-specific models. We validated this approach using non-invasive brain recordings from four linguistically diverse datasets covering English, Chinese, French and Dutch, demonstrating that UniDecoder successfully reconstructed semantically coherent sentences across all tested languages. Unlike the study by Silva et al., which leveraged shared articulatory representations to decode bilingual speech from invasive recordings \cite{RN15}, our approach employs semantic-level representations and non-invasive recordings to support multilingual decoding. Although the present study focuses on four languages, the PMM’s coverage of over 200 languages \cite{RN70} suggests that UniDecoder could be readily adapted to enable inclusive, multilingual BCI applications on a global scale.

UniDecoder leverages the unified semantic representation space to extend brain decoding capabilities to both multimodal and multi-subject settings. By projecting recordings from different modalities and individuals into the same semantic space, UniDecoder facilitates seamless integration of diverse data sources. In our experiments, this approach was validated using fMRI and MEG recordings, demonstrating that combining modalities enhances decoding performance compared to using any single modality. Although the validation focused on these two modalities, the framework is inherently modality-agnostic, as all recordings are projected into the same semantic representation space where fusion is performed \cite{us+8}. Such a method supports the integration of any combination of neuroimaging modalities, enabling the incorporation of complementary spatial and temporal information to improve decoding robustness and broaden the applicability of brain decoding technologies. Similarly, the use of a unified semantic space inherently supports generalization across individuals, as the mapping relies on language-agnostic semantic features rather than subject-specific neural patterns. To further address inter-subject variability and enhance alignment within this shared space, UniDecoder incorporates ISH to align brain recordings from different participants. This strategy enables effective multi-subject decoding while preserving individual characteristics, facilitating more practical and scalable BCI applications across diverse user populations.

Benefiting from the unified semantic space of UniDecoder, combined with its multilingual and multi-subject capabilities, the framework demonstrates robust decoding performance for resource-constrained subjects and supports extension to cross-lingual scenarios, contributing to enhanced lingual fairness. By leveraging the unified semantic space, UniDecoder enables the integration of data from other individuals to enhance semantic mapping for subjects with limited data availability. While this capability was observed in same-language settings, more importantly, our experiments showed that incorporating data from high-resource languages effectively improved decoding performance in low-resource language scenarios, highlighting the framework’s ability to support cross-lingual decoding. The design of UniDecoder, which integrates multilingual and multi-subject brain recordings within a unified semantic space, provides a conceptual basis for future expansion toward distributed learning frameworks. This approach shares similarities with federated learning strategies widely adopted in medical imaging domains \cite{RN75,us+5} and could support collaborative model development across decentralized neurodata, thereby lowering data and technical barriers and facilitating the broader adoption of BCI technologies \cite{RN76}.

Several critical challenges and limitations emerged from this study. First, the analysis revealed substantially lower decoding performance for Dutch compared to other languages. This may be related to the behavior of the PMM tokenizer currently used, which produces a higher degree of token fragmentation for Dutch, where many words are segmented into multiple subword tokens due to limited vocabulary coverage in the tokenizer. For example, sentences with identical semantic content require 24, 23, and 22 tokens in English, Chinese, and French, respectively, while Dutch necessitates 39 tokens (Extended Data Fig. 2). This increased token density complicates decoding by requiring the model to reconstruct a larger number of tokens from the same amount of brain recordings. A potential solution to mitigate this challenge would be to refine the tokenizer component of the PMM for Dutch by improving vocabulary coverage, thereby reducing unnecessary subword segmentation and enhancing decoding efficiency. Similar issues may also arise when applying the current PMM to other low-resource languages with complex morphology or compounding structures, highlighting a broader challenge that warrants further attention.

A further limitation concerns the observation that mismatches between the decoding language and the stimulus language resulted in notable degradation of reconstruction accuracy (Fig. \ref{fig:3}(b,d)). This phenomenon may be associated with the current word-level decoding strategy adopted in UniDecoder, which first maps brain recordings into a unified semantic representation at the word level before reconstructing text in the target language. Under this strategy, structural differences between languages, including word order and syntax, introduce challenges for accurate cross-language decoding. Although the current framework can correctly identify the source language within the four tested languages, such identification may become difficult when scaling to a broader set of languages, potentially leading to incorrect cross-language decoding. Addressing these issues will require advancing the decoding process toward sentence-level comprehension, enabling the extraction of language-independent semantic representations from brain recordings before generating expressions in the target language. Transitioning to sentence-level semantic reconstruction is expected to improve cross-language decoding performance.

A final and critical limitation relates to the precision of linguistic reconstruction achieved by UniDecoder. While the current framework enables reliable reconstruction of semantic-level content, it remains confined to approximate semantic representations, without reconstructing precise word-level output. This limitation reflects that the current decoding framework focuses on mapping brain activity to high-level semantic embeddings. However, it does not include the linguistic features needed for accurate word-level reconstruction. Advancing toward more fine-grained decoding would require integrating complementary linguistic features, such as phonological, acoustic, and speech motor representations, to disambiguate lexical units and capture speaker-specific nuances. Supporting this direction, recent work has demonstrated that unified acoustic-to-speech-to-language frameworks can better align with the hierarchical processing of natural speech and language in the brain, offering a promising computational path for enhancing decoding precision \cite{RN68}.

\section*{Methods}\label{sec4}
\subsection*{Problem formalization}
We aim to reconstruct natural language text that captures the meaning of speech heard by the subject. Each stimulus sentence is represented as a word sequence \( W = (w_1, \dots, w_L) \), which is tokenized by a PMM into a sequence of \( N \) tokens \( \mathcal{S} = (s_1, \dots, s_N) \). The PMM computes semantic embeddings for all tokens, forming a matrix \( \mathbf{Y} \in \mathbb{R}^{N \times d} \), where \( d \) is the embedding dimension determined by the number of parameters in the PMM. This matrix constitutes the unified semantic representation corresponding to the speech stimulus.

Simultaneously, brain responses evoked by the stimulus are recorded as a neural signal matrix \( \mathbf{X} \in \mathbb{R}^{t \times c} \), where \( t \) and \( c \) denote the number of time points and spatial channels, respectively. These dimensions vary across neuroimaging modalities and experimental configurations. For example, fMRI typically exhibits higher spatial resolution with lower temporal sampling, whereas EEG and MEG provide higher temporal resolution with fewer spatial channels \cite{RN60,RN61}.

The brain recordings \( \mathbf{X} \) are mapped into the unified semantic space, yielding a predicted semantic matrix \( \hat{\mathbf{Y}} \in \mathbb{R}^{N \times d} \) aligned with \( \mathbf{Y} \). From \( \hat{\mathbf{Y}} \), a decoded token sequence \( \tilde{\mathcal{S}} = (\tilde{s}_1, \dots, \tilde{s}_N) \) is generated using a semantic consistency-guided beam search algorithm. This sequence is then detokenized by the PMM into the final decoded word sequence \( \tilde{W} = (\tilde{w}_1, \dots, \tilde{w}_{\tilde{L}}) \), which represents the output of the decoding process.

\subsection*{Datasets}
The models were evaluated across four distinct datasets comprising 159 participants in total, all of which were approved by the relevant ethics committees and are publicly available for foundational research purposes. Key characteristics of the datasets are summarized in Extended Data Table 1. In the SMN4Lang dataset, 12 native Mandarin speakers listened to Mandarin news broadcasts while fMRI and MEG data were recorded \cite{RN10}. This study received approval from the Institutional Review Board of Peking University. In the LPPC-fMRI dataset, English, Chinese, and French versions of \textit{The Little Prince} were presented to 49 English-speaking, 35 Chinese-speaking, and 28 French-speaking healthy adults, with corresponding fMRI data collected \cite{RN8}. Ethical approval for this research was granted by the ethics committees of Cornell University, Jiangsu Normal University, and the French regional biomedical research ethics committee. In the Broderick's dataset, 19 English-speaking participants listened to excerpts from \textit{The Old Man and the Sea} while EEG data were recorded \cite{RN32}. This study was approved by the Ethics Committee of the School of Psychology and the Department of Health Sciences at Trinity College Dublin. For the SparrKULee dataset, we selected 16 native Dutch speakers from a larger cohort of 85 Dutch/Flemish-speaking healthy adults who listened to an audiobook while EEG data were recorded \cite{RN11}. This study was approved by the KU Leuven Medical Ethics Committee.

\subsection*{Data preprocessing}
fMRI data were processed using the ABCD-BIDS pipeline \cite{RN40} (ABCD BIDS Community Collection; NDA Collection 3165), which extends the Human Connectome Project pipeline \cite{RN29}. The processing workflow comprised six sequential stages: 1) PreFreeSurfer for denoising and spatial registration; 2) FreeSurfer for brain segmentation and cortical surface reconstruction; 3) PostFreeSurfer for CIFTI file generation; 4) fMRIVolume for motion and distortion corrections; 5) fMRISurface for mapping to standard CIFTI grayordinates in fs\_LR\_32k surface format; and 6) DCANBOLD processing for nuisance regression and motion censoring. Subsequently, we parcellated the brain according to the Destrieux atlas \cite{RN43}, selecting 26 language-processing regions of interest \cite{RN42}, including the precuneus, angular gyrus, temporal gyri (inferior, middle, and superior), and frontal gyri (inferior, middle, and superior), encompassing approximately 13,000 voxels for stimulus reconstruction, as shown in Fig. \ref{fig:1}d. To compensate for hemodynamic delay, we applied a 4-TR lag to align neural responses with stimulus presentation during decoding.

MEG and EEG data were preprocessed using independent component analysis for artifact removal. MEG recordings were additionally subjected to temporal signal space separation to eliminate magnetic interference artifacts. Both modalities underwent bandpass filtering (0.1-40 Hz), downsampling to 200 Hz, and z-score normalization with values exceeding 20 standard deviations clamped. Neural activity was extracted in 1-second epochs (±0.5s) centered on word onsets for subsequent decoding analyses.

For semantic processing of stimuli, we first converted auditory input to text using the Whisper model \cite{RN45}. We then applied PMM to extract semantic features from these texts. For this purpose, we selected the Bloom-1.1B model \cite{RN27} given its support for 46 languages and robust semantic representation capabilities. To enhance task-specific performance, we further refined the model using LLaMA-Factory \cite{llamafactory} on news and story-related corpora. Semantic features were derived from the 20th layer embeddings with a 15-token context window per segment to balance representational quality with computational constraints. These semantic features served as the target representations for our neural decoding framework.

\subsection*{UniDecoder}
\textbf{Self-adaptive semantic mapper.} The SSM serves as the core input module of UniDecoder, mapping brain recordings from diverse language stimuli, subjects, and neuroimaging modalities into a unified semantic space defined by a PMM. Brain recordings are first processed by neural encoders tailored to each modality. For fMRI, which provides high spatial but limited temporal resolution, Lanczos interpolation \cite{RN42} aligns the recordings with word-level stimulus timings, enabling voxel-wise mapping to semantic features. For MEG and EEG, which offer high temporal resolution, recordings from all channels within a ±1s window around each word stimulus are used to extract semantic representations. To enhance generalization, the fMRI encoder is pre-trained on the UK Biobank dataset \cite{RN67}, and the EEG encoder is initialized with a pre-trained EEGPT model \cite{EEGPT}. The encoded representations are then passed through an ISH module to reduce subject-specific variance, followed by a residual network that maps them into the unified semantic space. Finally, feature-level fusion is performed within this space to integrate the semantic representations across modalities.

\vspace{1em}

\noindent \textbf{Inter-subject Harmonization.} Inter-individual variability in brain responses presents a key challenge for building a multi-subject universal brain decoder---a single model capable of decoding brain recordings from multiple participants \cite{RN7}. To address this, UniDecoder incorporates the ISH module, which aligns neural representations from different subjects within the unified semantic space. Let \( \mathcal{U} \) denote the set of subjects, and for each \( u \in \mathcal{U} \), a subject-specific transformation matrix \( M_u \in \mathbb{R}^{d \times d} \) is applied to the output of the neural encoder. This transformation accounts for individual-specific variability and facilitates cross-subject alignment by projecting subject-dependent representations into a shared semantic structure. All other module parameters within UniDecoder are shared across \( \mathcal{U} \), enabling the framework to maintain a single model while flexibly adapting to multiple subjects. This design improves decoding accuracy in multi-subject settings without requiring trained models for individual participants.

\vspace{1em}

\noindent\textbf{Multimodal fusion.} To achieve more accurate brain decoding, we perform feature-level fusion of multimodal neuroimaging recordings within the unified semantic space. The unified semantic representations are obtained through SSM, which maps both fMRI and MEG signals into this space. Within the aligned space, modality-specific features are combined through weighted averaging to enhance decoding performance. fMRI provides high spatial resolution, while MEG offers high temporal resolution. This fusion strategy leverages the complementary strengths of each modality to improve the semantic precision of the reconstructed representations. Let \( M \) denote the total number of neuroimaging modalities, and let \( m \in \{1, \dots, M\} \) index the modality. The fused semantic representation is computed as:
\begin{equation}
\hat{\mathbf{Y}} = \sum\limits_{m = 1}^M \alpha_m \cdot \mathbf{F}_m, \label{eq1}
\end{equation}

\noindent
where \(\alpha_m\) denotes the fusion weight for the \(m\)-th modality, and \(\mathbf{F}_m \in \mathbb{R}^{N \times d}\) is its corresponding representation in the unified semantic space obtained via SSM. The fused output \(\hat{\mathbf{Y}} \in \mathbb{R}^{N \times d}\) serves as the final semantic representation for decoding.

\vspace{1em}
\noindent\textbf{Loss function.}
To map brain recordings into the unified semantic space defined by the PMM, we design a composite loss function that aligns brain-derived semantic representations with those extracted from stimulus texts. The total loss comprises three components: a directional cosine similarity loss \( \mathcal{L}_{\text{CS}} \), a token-level mean squared error (MSE) loss \( \mathcal{L}_{\text{MSE}} \), and a CLIP contrastive loss \( \mathcal{L}_{\text{CLIP}} \) \cite{CLIP}. Our composite loss function is defined as:
\begin{equation}
\begin{split}
\mathcal{L}_{\text{total}}(\mathbf{Y}, \hat{\mathbf{Y}}) =\,
& \beta_1 \cdot \mathcal{L}_{\text{CS}}(\mathbf{Y}, \hat{\mathbf{Y}}) \\
& + \beta_2 \cdot \mathcal{L}_{\text{MSE}}(\mathbf{Y}, \hat{\mathbf{Y}}) \\
& + \beta_3 \cdot \mathcal{L}_{\text{CLIP}}(\mathbf{Y}, \hat{\mathbf{Y}}),
\end{split}
\label{eq2}
\end{equation}
\noindent
where \( \beta_1 = 0.4 \), \( \beta_2 = 0.3 \), and \( \beta_3 = 0.3 \) are hyperparameters that control the relative contribution of each loss term and are set based on empirical validation. The individual loss terms are computed as:
\begin{equation}
\mathcal{L}_{\text{CS}} = \frac{1}{N} \sum_{i=1}^{N} \left( 1 - \frac{\mathbf{y}_i \cdot \hat{\mathbf{y}}_i}{\|\mathbf{y}_i\| \, \|\hat{\mathbf{y}}_i\|} \right),
\label{eq3}
\end{equation}

\begin{equation}
\mathcal{L}_{\text{MSE}} = \frac{1}{N} \sum_{i=1}^{N} \left\| \mathbf{y}_i - \hat{\mathbf{y}}_i \right\|^2,
\label{eq4}
\end{equation}

\begin{equation}
\mathcal{L}_{\text{CLIP}} = -\frac{1}{N} \sum_{i=1}^N \log \frac{\exp\left( \mathbf{y}_i \cdot \hat{\mathbf{y}}_i \right)}{\sum_{j=1}^N \exp\left( \mathbf{y}_i \cdot \hat{\mathbf{y}}_j \right)},
\label{eq5}
\end{equation}

\noindent
where \( \mathbf{y}_i, \hat{\mathbf{y}}_i \in \mathbb{R}^d \) represent the \( i \)-th row vectors of the semantic embedding matrices \( \mathbf{Y} \) and \( \hat{\mathbf{Y}} \), respectively, and \( j \) indexes contrastive candidates in the denominator of Eq.~\ref{eq5}.

\vspace{1em}
\noindent\textbf{Semantic consistency beam search.}
After mapping brain recordings into the unified semantic space, natural language text is generated from the resulting semantic representation using a semantic consistency-based extension of standard beam search \cite{RN31} in combination with the PMM. During decoding, each candidate sequence is evaluated based on three criteria: the token-level probability from the PMM to ensure linguistic fluency, the MSE between the generated embedding and the unified representation to ensure word-level correspondence, and the cosine similarity between them to promote semantic alignment. At each decoding step, a beam of \( k \) candidate sequences is maintained, and the scoring function is defined as follows:
\begin{equation}
\begin{split}
\text{score}(\tilde{\mathcal{S}}) = \sum_{i=1}^{N} \biggl( &\log P(\tilde{s}_i \mid \tilde{s}_1, \ldots, \tilde{s}_{i-1}) \\
&- \lambda_1 \cdot \| \hat{\mathbf{y}}_i - \tilde{\mathbf{y}}_i \|^2 \\
&+ \lambda_2 \cdot \frac{ \hat{\mathbf{y}}_i \cdot \tilde{\mathbf{y}}_i }{ \| \hat{\mathbf{y}}_i \| \, \| \tilde{\mathbf{y}}_i \| } \biggr),
\end{split}
\label{eq:beam_score}
\end{equation}

\noindent
where \( P(\tilde{s}_i \mid \tilde{s}_1, \ldots, \tilde{s}_{i-1}) \) is the token-level probability assigned by the PMM during generation. The vector \( \tilde{\mathbf{y}}_i \in \mathbb{R}^d \) denotes the semantic embedding of candidate token \( \tilde{s}_i \) extracted by the PMM, and \( \hat{\mathbf{y}}_i \in \mathbb{R}^d \) denotes the predicted semantic embedding of the \( i \)-th token decoded from brain recordings. The weighting parameters \( \lambda_1 = 0.3 \) and \( \lambda_2 = 0.7 \) are tunable hyperparameters used to balance semantic alignment and linguistic fluency. The final decoded word sequence is denoted as \( \tilde{W} \), obtained by detokenizing \( \tilde{\mathcal{S}} \) using the tokenizer of the PMM.

The beam search process iteratively expands and evaluates candidates, maintaining only the top-$k$ scoring sequences. Unlike conventional approaches that rely solely on language model probabilities, the decoding procedure is explicitly guided toward sequences that preserve the semantic content of the brain recordings. To promote diversity among candidates, the number of continuations from each hypothesis is limited. A beam width of 15 was adopted to increase exploration breadth and improve the semantic variability of generated hypotheses.  This semantic-guided decoding framework encourages the output text to remain consistent with the meanings represented in the brain recordings while preserving naturalistic language structure.
\vspace{1em}

\noindent\textbf{Multilingual decoding.} To support multilingual brain decoding, UniDecoder maps brain recordings into a unified semantic space defined by a PMM. The semantic space was constructed from the embedding representations of the PMM, which encodes semantic structures shared across languages through training on large-scale multilingual corpora. Brain recordings acquired under different language conditions were projected into this space using the SSM. In cases where the stimulus language was unknown, a language recognition module was introduced. This module determines the most probable source language from the unified semantic representation, and the predicted label is used to guide language-specific text generation. This design enables language-conditioned decoding without requiring prior knowledge of the stimulus language.

\subsection*{SHAP analysis}
To investigate the neural basis of semantic processing, we sought to quantify how different brain regions contribute to semantic decoding. We employed the SHAP framework \cite{SHAP}, which provides a unified approach for model interpretation based on cooperative game theory. SHAP values determine feature importance by measuring their marginal contributions across all possible feature combinations, thereby ensuring fair attribution of model outputs to input features. This framework is particularly suitable for analyzing complex neural architectures like our adaptive multimodal mapper.
For our analysis of fMRI recordings, we first computed voxel-wise SHAP values to quantify each voxel's contribution to the predicted semantic vectors generated by our decoder. To obtain a more comprehensive understanding of regional involvement in semantic processing, these voxel-level contributions were then aggregated according to the Destrieux atlas \cite{RN43}, providing region-wise SHAP values. The magnitude of these aggregated SHAP values indicates the strength of each region's influence on the semantic decoding process, with higher absolute values suggesting greater contributions to semantic representation.

\subsection*{Language similarity metrics}
Four distinct evaluation metrics were employed to compare decoded word sequences with reference counterparts. WER quantifies discrepancies by measuring the number of edit operations required to transform predicted sequences into references, with lower values indicating greater similarity. BLEU-1 \cite{RN53} evaluates the proportion of word matches between generated and reference texts. METEOR \cite{RN54} assesses similarity by integrating precision, recall, and synonym matching. BGEScore is a semantic similarity metric proposed in this study, derived from the multilingual BGE-M3 embedding model \cite{RN26}. It computes the cosine similarity between sentence-level dense embeddings of the generated and reference texts. Unlike traditional lexical metrics, BGEScore projects texts into a shared semantic space, enabling direct, language-agnostic comparisons of semantic similarity across languages.

To enable direct comparison across evaluation metrics with distinct scales and distributions, we transformed all metric scores into z-scores relative to the mean and variance of scores computed from randomly generated outputs. This yields the normalized similarity $ \mathcal{Z}_{\mathrm{sim}} $, which quantifies how much each decoded output improves over randomly generated text from perturbed semantic representations. The z-score normalized similarity is computed as:
\begin{equation}
\mathcal{Z}_{\mathrm{sim}} = \frac{ \mathrm{sim}(W, \hat{W}) - \mu }{ \sigma },
\label{eq:7}
\end{equation}
\noindent
where \( \mathrm{sim}(\cdot, \cdot) \) denotes the similarity between two sentences computed using a predefined metric (e.g., WER, BLEU-1, METEOR, or BGEScore), and \( \mu \) and \( \sigma \) represent the mean and standard deviation of similarity scores computed over 200 generations in which brain-derived semantic representations were replaced with randomly sampled vectors.

\bibliography{sn-bibliography}

\section*{Competing interests}\label{sec10}
The authors declare no competing interests.

\setcounter{table}{0}
\renewcommand{\tablename}{}
\renewcommand{\thetable}{Extended Data Table \arabic{table}}

\setcounter{figure}{0}
\renewcommand{\figurename}{Extended Data Fig.}  
\renewcommand{\thefigure}{\arabic{figure}}  

\begin{table*}[t!]
\centering
\caption{Datasets}

\begin{tabular}{@{}llllll@{}}
\toprule
\textbf{Dataset}        & \textbf{Language} & \textbf{Modal}  & \textbf{Participants} & \textbf{Participant Time} & \textbf{Total Time} \\ \midrule
SMN4Lang \cite{RN10}      & Chinese           & fMRI/MEG        & 12                     & 6 h                       & 72 h                \\
LPPC-fMRI (EN) \cite{RN8}    & English           & fMRI            & 49                     & 1.5 h                     & 73.5 h              \\
LPPC-fMRI (CN)                         & Chinese           & fMRI            & 35                     & 1.5 h                     & 52.5 h              \\
LPPC-fMRI (FR)                         & French            & fMRI            & 28                     & 1.5 h                     & 42 h                \\
Broderick2018 \cite{RN32} & English           & EEG             & 19                     & 1 h                       & 19 h                \\
SparrKULee \cite{RN11}   & Dutch             & EEG             & 16                     & 2 h                       & 32 h               \\ \bottomrule
\end{tabular}
\vbox{
\vspace{0.5em}
\noindent\parbox{\textwidth}{\small We analyzed four datasets, covering four languages (English, Chinese, French, and Dutch) and spanning three modalities (fMRI, EEG, and MEG). The table summarizes key statistics, including the number of participants, the average time per participant, and the total recording time.}
}
\label{tab:E1}
\end{table*}

\clearpage

\begin{figure*}
    \centering
    \includegraphics[width=\textwidth]{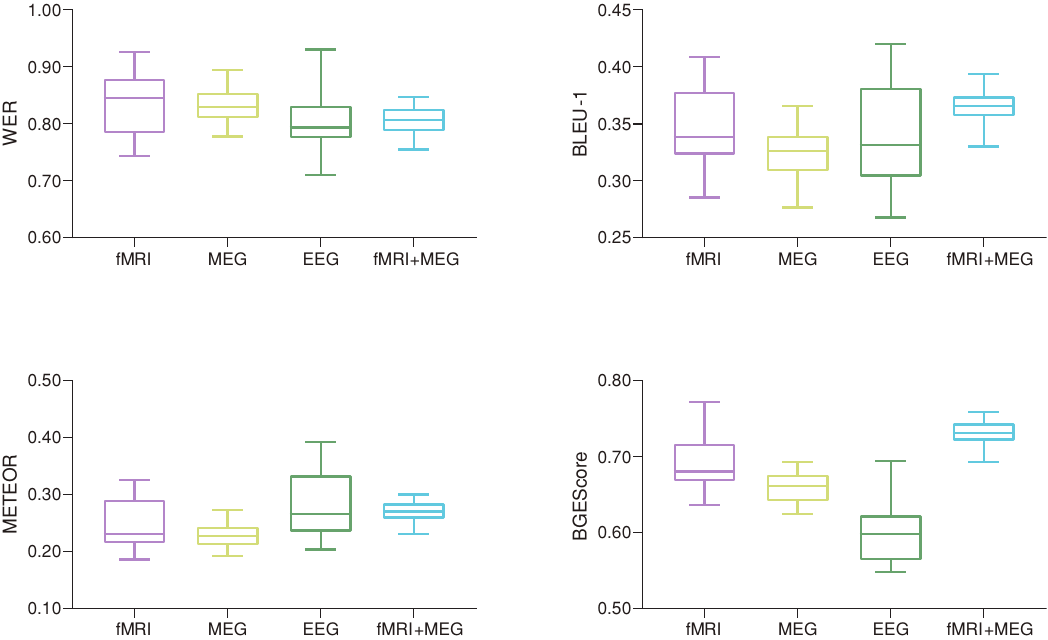}
    \caption{\textbf{Comparison of decoding performance across different modalities for the four datasets.} Box plots show the evaluation metrics: WER, BLEU-1, METEOR, and BGE Score for fMRI, MEG, EEG, and fMRI+MEG. The fusion of multiple modalities generally improves performance compared to individual modalities, with fMRI+MEG yielding better results across most metrics.} %
    \label{fig:ex1}
\end{figure*}

\clearpage

\begin{figure*}
    \centering
    \includegraphics[width=\textwidth]{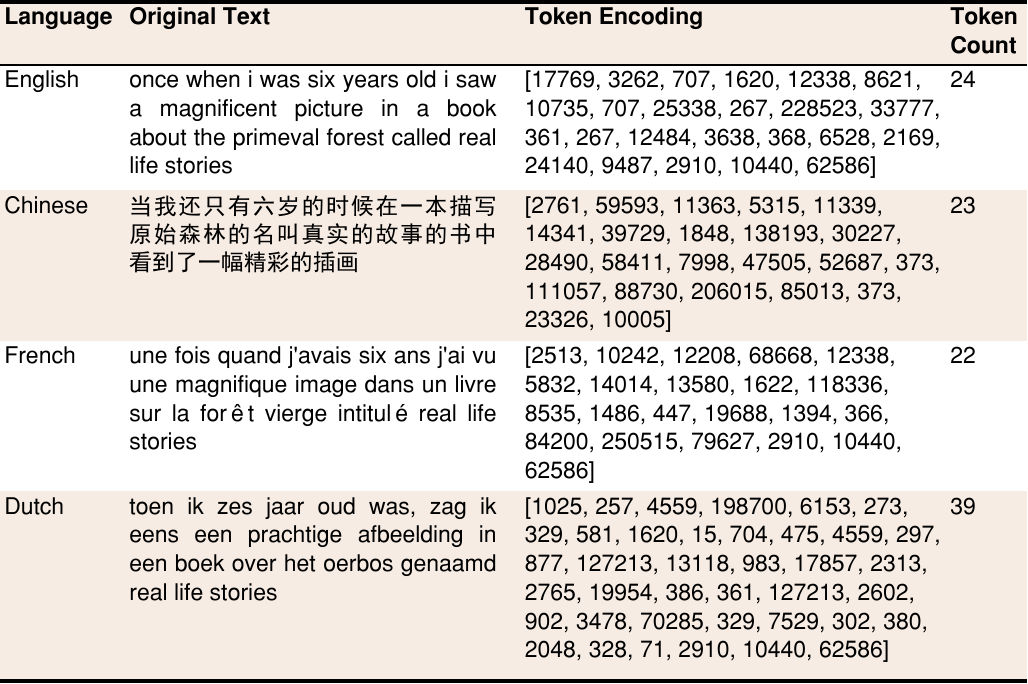}
    \caption{\textbf{Visualization of token encoding variation across semantically equivalent content in four languages.} Dutch displays substantially higher tokenization complexity compared to other languages, potentially increasing decoding difficulty for brain-based language processing models. Such tokenization differential highlights a critical challenge in multilingual neural decoding approaches, where languages with higher token counts may require more sophisticated computational resources and algorithms to achieve comparable decoding accuracy.} %
    \label{fig:ex2}
\end{figure*}

\end{document}